# Ultrafast All Optical Magnetization Control for Broadband Terahertz Spin Wave Generation

Saeedeh Mokarian Zanjani[1], Mehmet C. Onbaşlı[1,2,*]

[1]Graduate School of Materials Science and Engineering, Koç University, Sarıyer, 34450 Istanbul, Turkey.
[2] Department of Electrical and Electronics Engineering, Koç University, Sarıyer, 34450 Istanbul, Turkey.
* Corresponding Author: monbasli@ku.edu.tr

Terahertz spin waves could be generated on-demand via all-optical manipulation of magnetization by femtosecond laser pulse. Here, we present an energy balance model, which explains the energy transfer rates from laser pulse to electron bath coupled with phonon, spin, and magnetization of five different magnetic metallic thin films: Iron, Cobalt, Nickel, Gadolinium and Ni$_2$MnSn Heusler alloy. Two types of transient magnetization dynamics emerge in metallic magnetic thin films based on their Curie temperatures ($T_C$): type I (Fe, Co, and Ni with $T_C$ > room temperature, RT) and type II films (Gd and Ni$_2$MnSn with $T_C \approx$ RT). We study the effect of laser fluence and pulse width for single Gaussian laser pulses and the effect of metal film thickness on magnetization dynamics. Spectral dynamics show that broadband spin waves up to 24 THz could be generated by all-optical manipulation of magnetization in these nanofilms.

## I. INTRODUCTION

Terahertz (THz) spintronic systems have been emerging due to their new fundamental physics on the interactions of magnons, phonons and photons [1-9] as well as their applications in sensing, spectroscopy [10], and on-chip communications [11,12]. THz band contains many molecule-specific absorption peaks which could be used for highly-specific sensor designs. While electromagnetic sensing in the THz region using THz time-domain spectroscopy (TDS) has been demonstrated to have high specificity [13-18], spintronic THz spectroscopic systems could enable much higher sensitivities since THz spin waves are highly localized (sub-10 nm) compared to electromagnetic THz modes (240 µm) [10]. Thus, spintronic THz-TDS systems could be used for label-free single molecule detection.

In the design of such THz spintronic sensor systems, broadband THz spin waves could be generated by applying femtosecond (fs) laser pulses on metallic magnetic thin films [8, 9]. A detailed understanding of how the key material and laser parameters determine the generated THz spin wave spectra is essential for both sensing applications and for understanding phonon, magnon and photon interactions. In this study, we developed an energy balance model, which describes the coupled interaction between electron, phonon, and spin temperatures with normalized magnetization (extended M3TM) after illumination with fs Gaussian single laser pulse for five different metallic magnetic thin films: Iron (Fe), Cobalt (Co), Nickel (Ni), Gadolinium (Gd), and Ni$_2$MnSn Heusler alloy. We investigated the effect of the material properties of these metals and fs laser pulse characteristics (laser fluence and pulse width) on magnetization dynamics. Magnetization dynamics indicate that broadband THz spin waves could be generated inside the films due to exchange coupling. We provide the guidelines for obtaining a desired spin wave spectrum using the appropriate metallic film type, thickness, laser pulse width and fluence. By identifying the regimes in which $T_C$, phonon scattering, thermal mass and spin-flip scattering dominates, we explain the mechanisms that determine these spectra.

The interaction of fs laser pulse with Nickel (Ni) metallic magnetic materials was first explained using three temperature model (3TM) [19]. Since magnetization dynamics were not considered in this model, Landau-Liftshitz-Bloch (LLB) equation [20] was developed to model the laser-matter interaction. The LLB model explains the thermal effects of laser pulse on magnetization dynamics. Nevertheless, this model does not capture the thermal interaction of laser pulse with electron, phonon, and spin temperatures and their interactions. This model also depends on temperature dependent damping parameter, which was not measured for most materials. Koopmans et al. [21] developed a model considering the coupling of magnetization dynamics with electron and phonon temperatures and could extract the spin wave emission spectra; however, they neglected the coupling of spin temperature with that of phonon and electron in their model. Our extended M3TM model describes each of these effects for different pulse fluences, widths, film thicknesses and metal types. By capturing the heat exchange between electron, phonon and spin baths, their heat exchanges, the magnetization dynamics and their spectral characteristics, our model is the first to identify the conditions needed for broadband and tunable THz spin wave emission.

## II. ENERGY BALANCE MODEL AND MAGNETIZATION DYNAMICS

We developed an extended three-temperature model coupled with transient magnetization dynamics to capture the energy transfer rates between spin, phonon and electron thermal baths. These thermal baths are assumed to behave in the classical or semi-bulk regime (i.e. films are sufficiently thick with no quantum confinement effects, t > 20 nm). The energy balance model describes that the fs laser pulse injects energy into the coupled baths and the transient electron, phonon, and spin temperatures ($T_e$, $T_p$, $T_s$) and magnetization are described with the rate equations (1) to (4) [19, 21].

$$C_e \frac{dT_e}{dt} = -G_{ep}(T_e - T_p) - G_{es}(T_e - T_s) + P(t) \quad (1)$$

$$C_p \frac{dT_p}{dt} = -G_{ep}(T_p - T_e) - G_{ps}(T_p - T_s) \quad (2)$$

$$C_s \frac{dT_s}{dt} = -G_{es}(T_s - T_e) - G_{ps}(T_s - T_p) \quad (3)$$

$$\frac{dm}{dt} = Rm \frac{T_p}{T_C} \left(1 - m\coth\left(m\frac{T_C}{T_e}\right)\right) \quad (4)$$

In these equations, $T_e$, $T_p$, $T_s$, and $T_C$ are electron, phonon, spin, and Curie temperatures, respectively. R is the spin-flip ratio [21] and determines the kinetics of transient changes of $T_e$, $T_p$, $T_s$, and magnetization. The electron-phonon, electron-spin, and phonon-spin coupling constants $G_{ep}$, $G_{es}$, $G_{ps}$, follow a similar formalism described in ref. [21] and [22]. The heat capacities of phonon and spin are $C_p$ and $C_s$. The electron heat capacity is a temperature-dependent parameter and is defined as $C_e = \gamma T_e$, where $\gamma$ is a parameter that depends on free electron density and Fermi energy level [23] as used in ref. [21] as $C_p = 5 \gamma T_C$). Finally, P(t) represents the laser pulse power injected per unit volume and **m** is normalized magnetization defined as $|M_z|/M_s$. The material parameters and constants used in the numerical solutions to the equations (1) to (4) for Ni, Co, Fe, Gd, and Ni$_2$MnSn are shown in Supplementary Table 1.

## III. RESULTS AND DISCUSSION

Fig.1 (a) shows a schematic of THz spin wave generation inside the metallic magnetic thin film. As a fs laser pulse hits the metallic film, the film's magnetic moment undergoes precession and damped oscillations. These oscillations trigger THz spin waves over their neighbors due to exchange coupling.

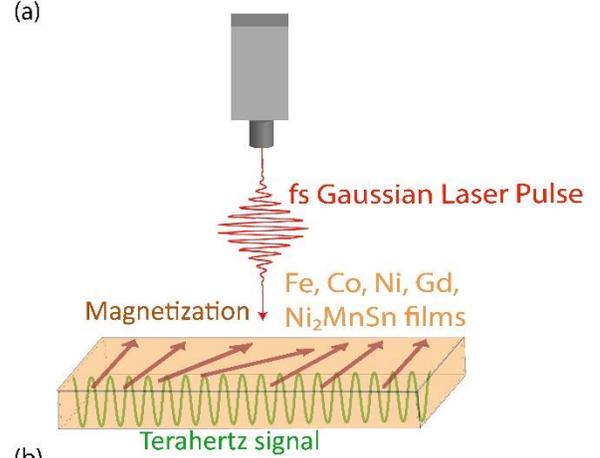

(a)

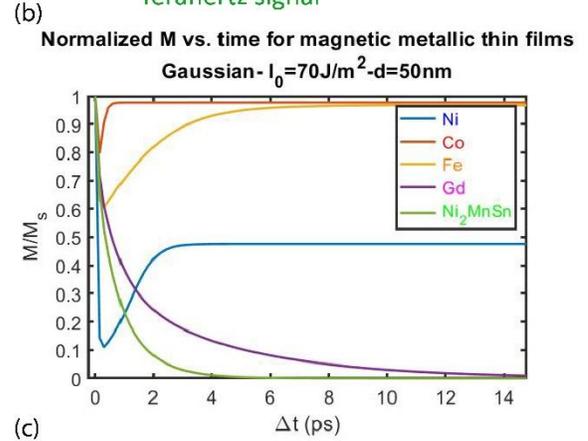

(b)

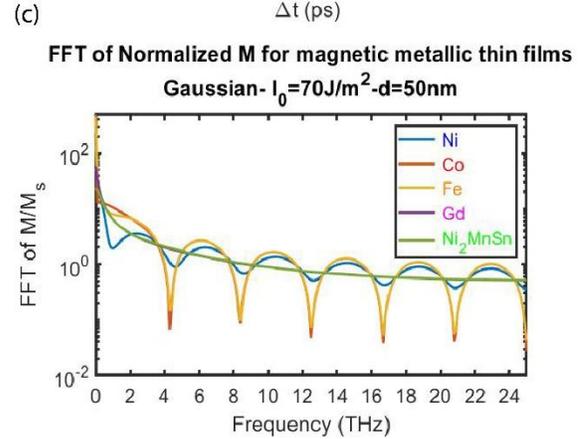

(c)

**FIG. 1. Effect of fs laser pulse on metallic magnetic thin films. (a)** THz spin wave generation in metallic magnetic thin films after illumination with a fs laser pulse. **(b)** Transient magnetization dynamics of 50 nm thick Fe, Co, Ni, Gd, and Ni$_2$MnSn Heusler alloy films illuminated with 100 fs laser pulse with 70 J·m$^{-2}$ fluence. **(c)** Exchange-coupled spin wave spectra generated after laser pulse (Fourier transform of magnetization in **b**).

Fig. 1(b) shows the calculated time-dependent relaxation of the magnetic moments of 50 nm thick Fe, Ni, Co, Gd, and Ni$_2$MnSn upon each receiving a 100 fs-long pulse with a moderate (70 J·m$^{-2}$) fluence. According to Fig.1 (b), magnetizations in Fe, Ni, Co (type I) reach the dip in less than 200 fs and reach steady state moments in 6, 3, and 0.5 ps, respectively.

We group Fe, Co, and Ni as type I since their Curie temperatures greatly exceed room temperature (Fe: 1043 K, Co: 1388 K, Ni: 627 K). Thus, these films never undergo thermal demagnetization after interacting with fs laser pulse. On the other hand, Gd and Ni$_2$MnSn (type II) with Curie temperatures near room temperature (Gd: 297 K, Ni$_2$MnSn: 319 K) undergo thermal demagnetization upon interacting with the fs laser pulse and reach zero magnetization within 15 and 6 ps, respectively. The laser-induced loss of spin angular momentum could be due to spin-flip scattering with phonons and hot electrons (electron-phonon/electron-electron interaction), known as Elliot-Yafet scattering [24, 25]. The spin wave spectra for these thin films are shown in Fig. 1(c) by calculating the fast Fourier transform (FFT) of magnetization in Fig. 1(b). Broadband spin waves could be generated using Fe, Ni, and Co magnetic thin films up to 24 THz. The intensity of the THz signal is higher for Fe and Co due to their higher Curie temperatures than that for Ni. In addition, the recovered fraction of magnetization after a few picoseconds is higher in Fe and Co (M/M$_s$ > 97%) compared to Ni (M/M$_s$ ~50%). The recovered fraction of magnetization increases with increasing the film thickness.

In the rest of this section, the effect of film thickness, fs laser pulse fluence and pulse width on the magnetization dynamics and the spin wave spectra in both type I and II films are investigated.

### 1. Thin films with type I dynamics (Fe, Co, Ni)

Type I dynamics is defined as ultrafast switching/decrease of magnetization in fs timescales, followed by slow recovery of magnetization in picosecond timescales. Hot electrons excite transient magnetization loss and a subsequent excitation of spin waves (magnons) due to exchange coupling. Magnons undergo inelastic scattering with phonons and electronic charges in the lattice. The lattice serves as a reservoir for absorbed, dissipated and scattered energy and causes evanescent decay of the generated spin waves. Magnetization recovery time is slightly longer than electron-lattice relaxation time due to the different coupling strengths among lattice, phonon and electron temperatures (i.e. strength of Hamiltonian terms for each quasiparticle). In our model, we do not consider the quantum nature of these baths or their Hamiltonians and lump their couplings into G$_{ep}$, G$_{es}$, G$_{ps}$ parameters. These parameters are known for the metallic magnetic thin films in our study [21].

Fig. 2(a) shows the magnetization dynamics of (i) 20 nm and (ii) 100 nm thick Fe films under illumination with a 100 fs Gaussian laser pulse.

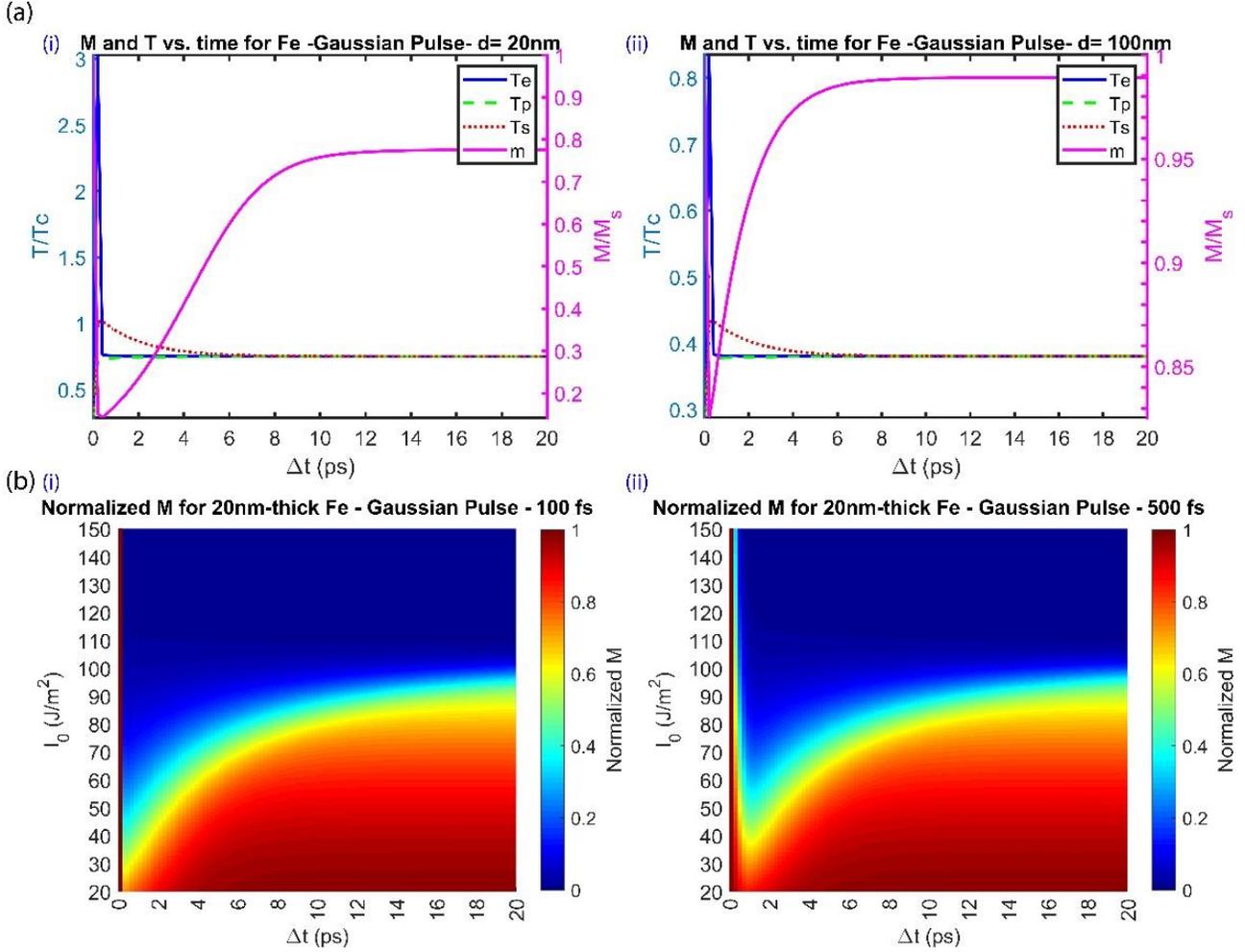

**FIG. 2. Effect of film thickness, fs laser fluence and pulse width on magnetization dynamics and lattice temperature of Fe thin film.** (a) Transient electron ($T_e$), phonon ($T_p$), spin ($T_s$) temperatures and magnetization (m=|$M_z$|/$M_s$) of (i) 20 nm and (ii) 100 nm thick Fe, illuminated with 100 fs laser pulse with 70 J·m$^{-2}$ fluence. (b) Fluence dependence of magnetization in 20 nm thick Fe film illuminated with (i) 100 fs and (ii) 500 fs laser pulse.

Following type I dynamics, magnetization of Fe starts switching around 300 fs for 20 nm film and slightly shorter than that for 100 nm Fe film. The perpendicular magnetization starts recovering and eventually reaches ~78% of saturation moment after 12 ps in 20 nm Fe film. The recovered magnetization fraction reaches 99% for 100 nm thick film after 8 ps. Due to the interaction of spins with the fs laser pulse, the electron, spin and lattice temperatures increase. Although the electron temperature undergoes a sudden increase to above Curie temperature ($T_C$) of Fe, due to electron's lower heat capacity ($C_e$), it drops quickly and the equilibrium lattice temperature ($T_p$) stays below $T_C$. Thus, both 20 and 100 nm Fe films retain their magnetization.

In Fig. 3(a), the magnetization dynamics of (i) 20 and (ii) 100 nm thick Co, are shown after illumination with Gaussian single pulse with 100 fs duration and 70 J·m$^{-2}$ fluence. Co magnetization (type I) decreases within 250 fs in 20 nm film and recovers back to 75% of its saturation moment in around 10 ps. Magnetic moment in 100 nm Co starts switching in ~160 fs and almost completely recovers (~99.5%) its saturation state after 800 fs. In this case, $T_p/T_C$ <1 so the film does not demagnetize completely.

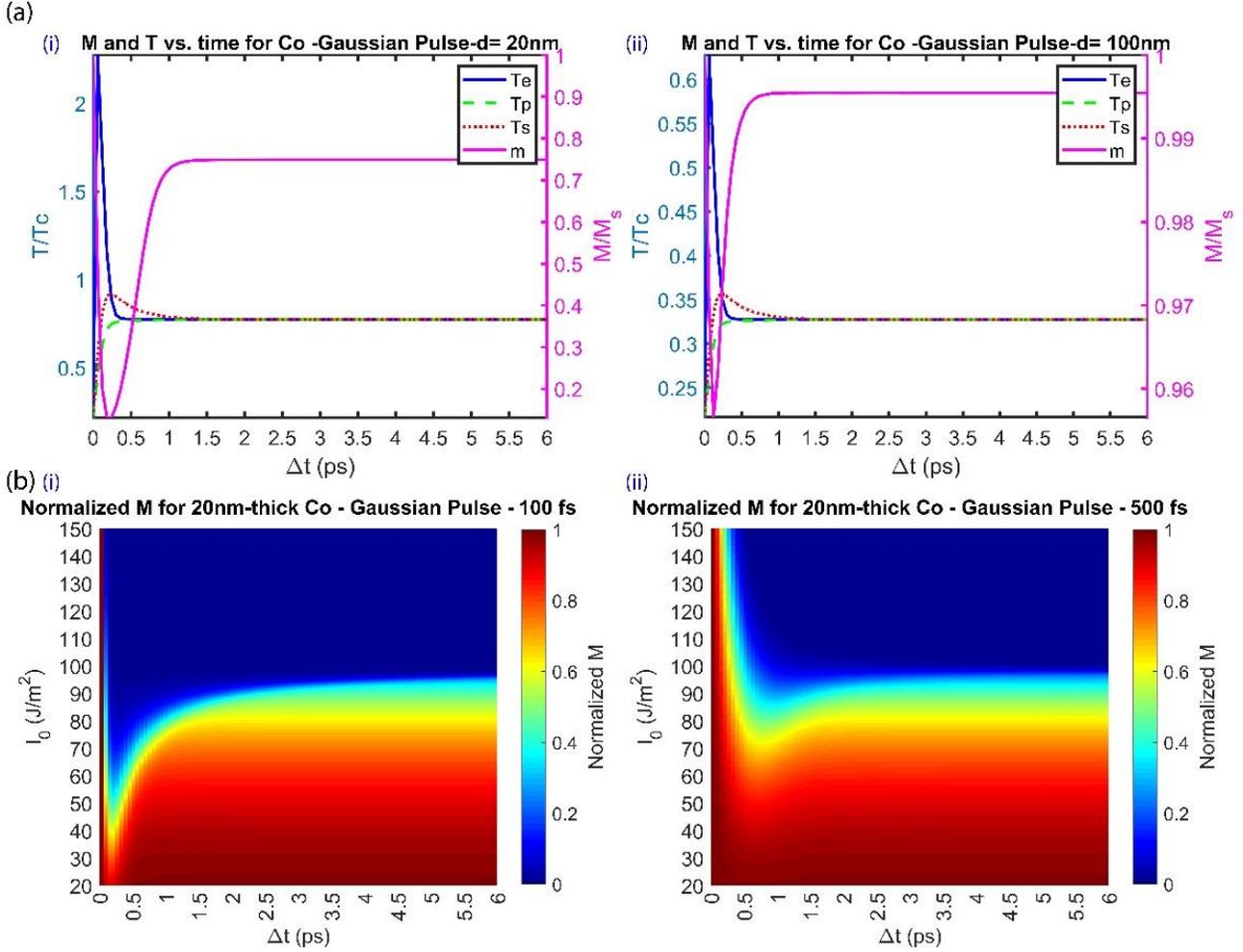

**FIG. 3. Effect of film thickness, fs laser fluence and pulse width on magnetization dynamics and lattice temperature of Co thin film.** (a) Transient electron ($T_e$), phonon ($T_p$), spin ($T_s$) temperatures, and normalized magnetization ($m=|M_z|/M_s$) of (i) 20 nm and (ii) 100 nm thick Co illuminated with 100 fs laser pulse with 70 J·m$^{-2}$ fluence. (b) Fluence dependence of normalized magnetization of 20 nm thick Co film illuminated with (i) 100 fs and (ii) 500 fs laser pulse.

In Fig. 4, the magnetization dynamics of (i) 20 and (ii) 100 nm thick Ni thin films upon illumination with 100 fs Gaussian pulses are shown. The 20 nm Ni film demagnetizes completely after 130 fs, since the lattice temperature ($T_p$) greatly exceeds nickel's Curie temperature. Nickel's $T_C$ (627 K) is lower than those of Fe (1043 K) and Co (1388 K); so 20 nm thick Ni film cannot retain its magnetization after interacting with the fs laser pulse. However, the thicker Ni film (100 nm) exhibits type I dynamic similar to Fe and Co films due to its larger heat capacity. The film's magnetic moment decreases in ~170 fs and recovers ~83% of its saturation moment within 1.2 ps. Increasing the film thickness in Ni decreases both switching and recovery times, but it increases the recovered fraction of magnetization. The fluence and pulse width dependence of magnetizations for 20 nm Fe (Fig. 2b), Co (Fig. 3b), Ni (Fig. 4b) films are shown. The magnetization dynamics have been calculated for fluences from 20 to 150 J·m$^{-2}$. Illuminated with both 100 fs (Fig. 2.b (i), Fig. 3. b (i), Fig. 4.b (i)) and 500 fs (Fig. 2.b (ii), Fig. 3. b (ii), Fig. 4.b (ii)), both magnetization switching and recovery times of 20 nm films increase with increasing laser fluence. The recovered magnetization fraction, however, decreases with increasing laser fluence. In addition, there is a threshold fluence above which the films are thermally demagnetized completely. This threshold fluence depends on the film type, thickness and pulse width. Increasing the pulse width increases the response time of the films, which is attributed to the larger energy injected and dissipated with the films. The recovered fraction of magnetization is higher in case of thicker films, since the higher thermal mass prevents the excess heat accumulation and magnetization disturbance in these films. The fluence dependence of magnetization dynamics of 100 nm films are presented in the Supplemental Material.

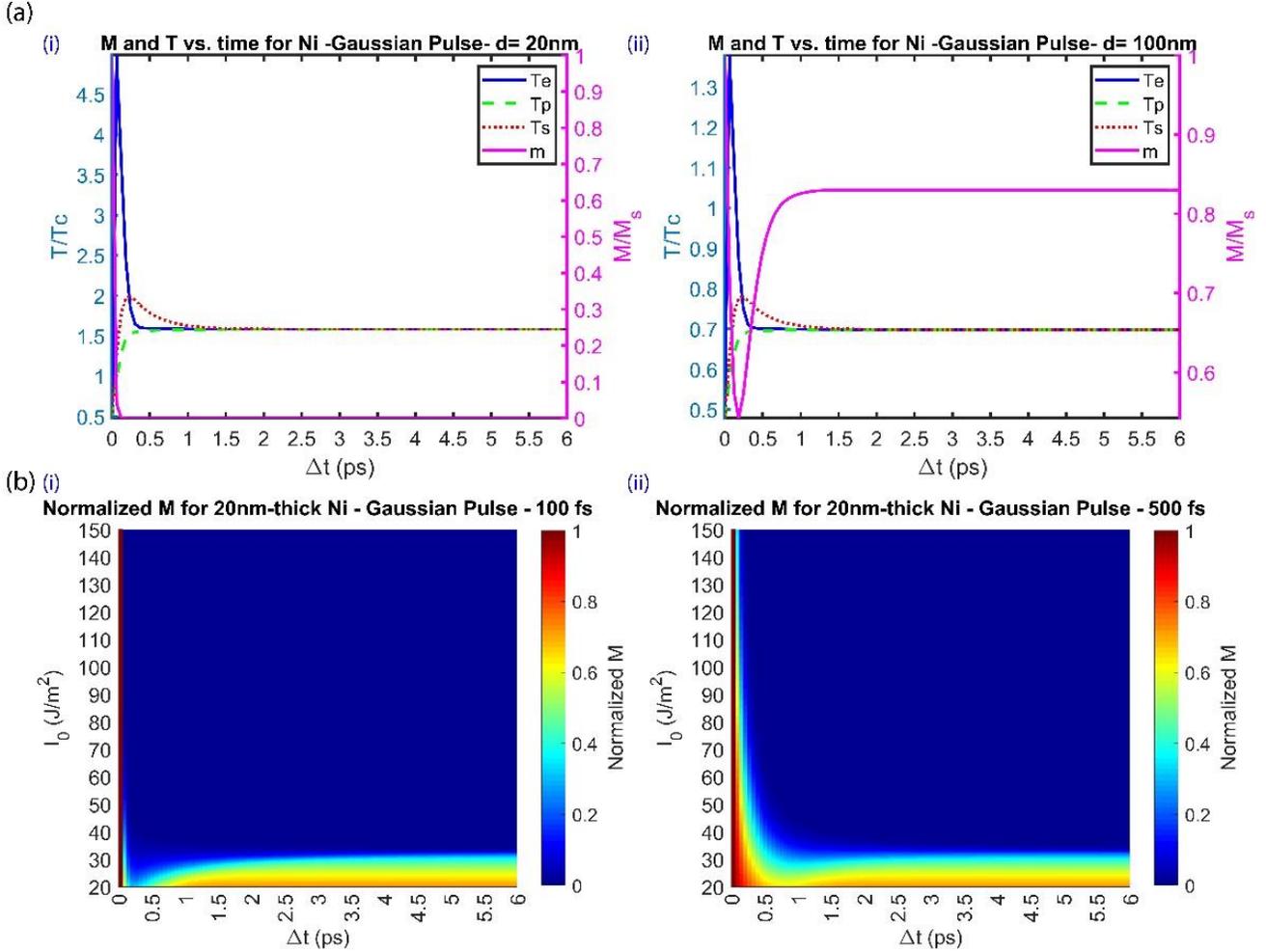

**FIG. 4. Effect of film thickness, fs laser fluence and pulse width on magnetization dynamics and lattice temperature of Ni thin film. a**(i) Transient electron ($T_e$), phonon ($T_p$), spin ($T_s$) temperatures, and normalized magnetization (m=|$M_z$|/$M_s$) of 20 nm and (ii) 100 nm thick Ni illuminated with 100 fs laser pulse with 70 J·m$^{-2}$ fluence . **b** Fluence dependence of normalized magnetization of (i) 20 nm thick Ni film illuminated with 100 fs and (ii) 500 fs laser pulse.

### 2. THz spin wave spectra generated in type I films

Fig. 5 shows the fast Fourier transform (FFT) results of the fluence dependence of temporal magnetization change for Fe, Co and Ni. As shown in Fig. 5, by illuminating the Fe, Co, and Ni films with single 100 fs Gaussian laser pulses each, spin waves up to 10 THz could be generated. The THz spin wave intensity is higher when fs laser fluence is lower. Increasing film thickness to 100 nm increases the THz spin wave intensity and the allowed threshold fluence for THz spin wave generation. Laser pulses with sub-picosecond widths have minimal effect on the bandwidth and the intensity of the THz spin wave signal (see the spectra of the films under illumination of 500 fs laser pulse in Supplemental Material). From the spectra of Fe and Co, one can infer that the THz spin wave intensity is slightly higher compared to Ni, since these two metals have higher $T_C$ and recovered magnetization fractions (> 95%).

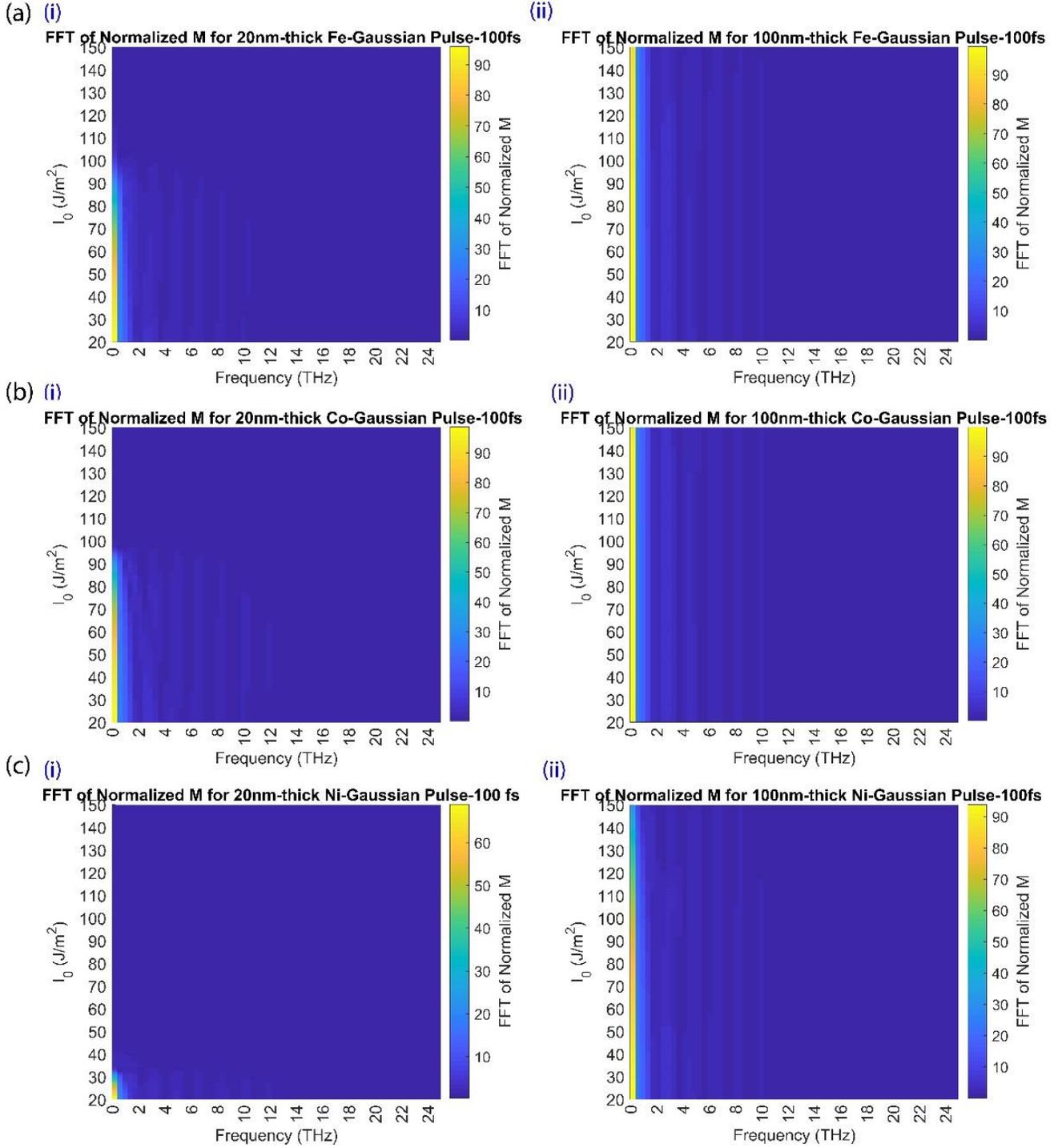

**FIG. 5. THz spin wave generation in type I metallic magnetic thin films.** Effect of laser fluence on the bandwidth and intensity of THz spin wave generated in **a**(i) 20nm thick, **a**(ii) 100 nm thick Fe, **b**(i) 20nm thick, **a** (ii) 100 nm thick Co, and **a**(i) 20nm thick, **a**(ii) 100 nm thick Ni thin films under illumination of 100 fs laser pulse.

### 3. Thin films with type II dynamics (Gd, Ni$_2$MnSn)

Type II magnetization dynamics arise when the films undergo thermal demagnetization (magnetic moment decays to zero in tens of picoseconds). Fig. 6 and 7 show the magnetization dynamics of type II films, Gd and Ni$_2$MnSn. Fig. 6a (i) and (ii) show the temporal change of $T_e$, $T_p$, $T_s$, and the magnetization of 20 nm and 100 nm Gd thin films after interacting with a 100 fs Gaussian single laser pulse of 70 J·m$^{-2}$ fluence. The spin-flip ratio of Gd is small (R= 0.092×10$^{12}$ s$^{-1}$) compared to the type I metallic magnetic thin films; so the magnetization vanishes slowly after 2.5 ps in 20 nm film and 50 ps in 100 nm Gd film. Since the Curie temperature of Gd is near room temperature (297 K), the equilibrium electron, spin and the lattice temperatures exceed the Curie temperature and the film is thermally demagnetized completely with the laser pulse. The equilibrium lattice temperature of 20 nm thick Gd film (Fig. 6.a (i)) is higher compared to the 100 nm thick film (Fig. 6.a (ii)) due to the lower thermal mass of the thinner film. Similarly, Ni$_2$MnSn also undergoes thermal demagnetization and its magnetization decays faster than Gd due to the alloy's larger spin-flip ratio (R=0.1×10$^{12}$s$^{-1}$). Demagnetization times of 20 nm and 100 nm thick Ni$_2$MnSn films are 1 ps and 35 ps, respectively, showing that a larger thermal mass delays thermal equilibration. Fig. 6(b) and 7(b) show the dependence of magnetization dynamics to the laser pulse fluence and duration.

Similar to type I dynamics, increasing the laser fluence and pulse duration, increases the response time in both type II films. On the other hand, excitation of the films with smaller laser fluences keeps the films magnetized for longer times. Unlike type I films, the response time increases in type II by increasing the film thickness. According to Fig. 6(b) and Fig. 7(b), increasing the incoming laser pulse width increases the response time.

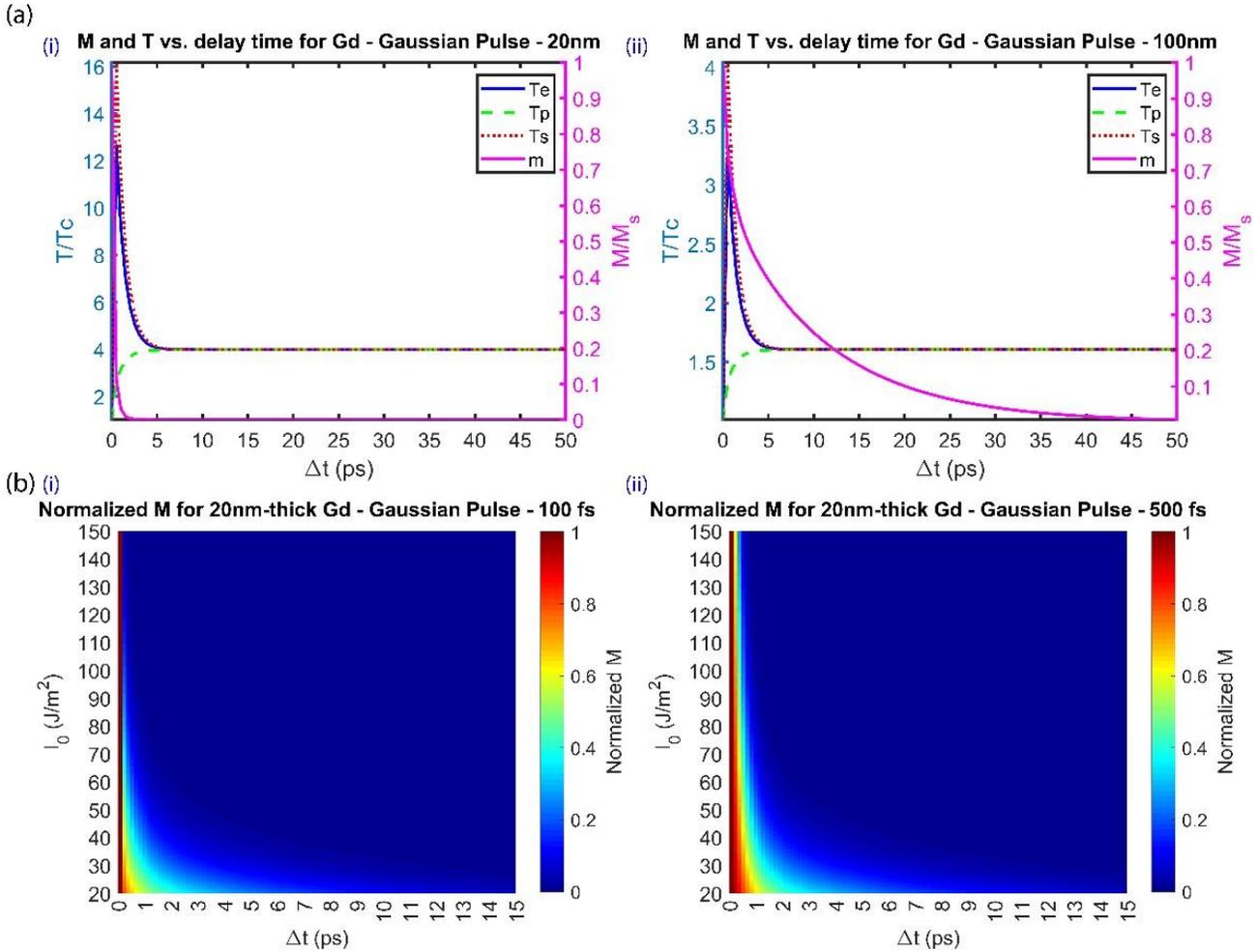

**FIG. 6. Effect of film thickness, fs laser fluence and pulse width on magnetization dynamics and lattice temperature of Gd magnetic thin film.** (a) transient electron (T$_e$), phonon (T$_p$), spin (T$_s$) temperatures, and normalized magnetization (m=|M$_z$|/M$_s$) of (i) 20 nm and (ii) 100 nm thick Gd illuminated with 100 fs laser pulse with 70 J·m$^{-2}$ fluence. (b) Fluence

dependence of normalized magnetization of 20 nm thick Gd film illuminated with (i) 100 fs and (ii) 500 fs Gaussian laser pulse.

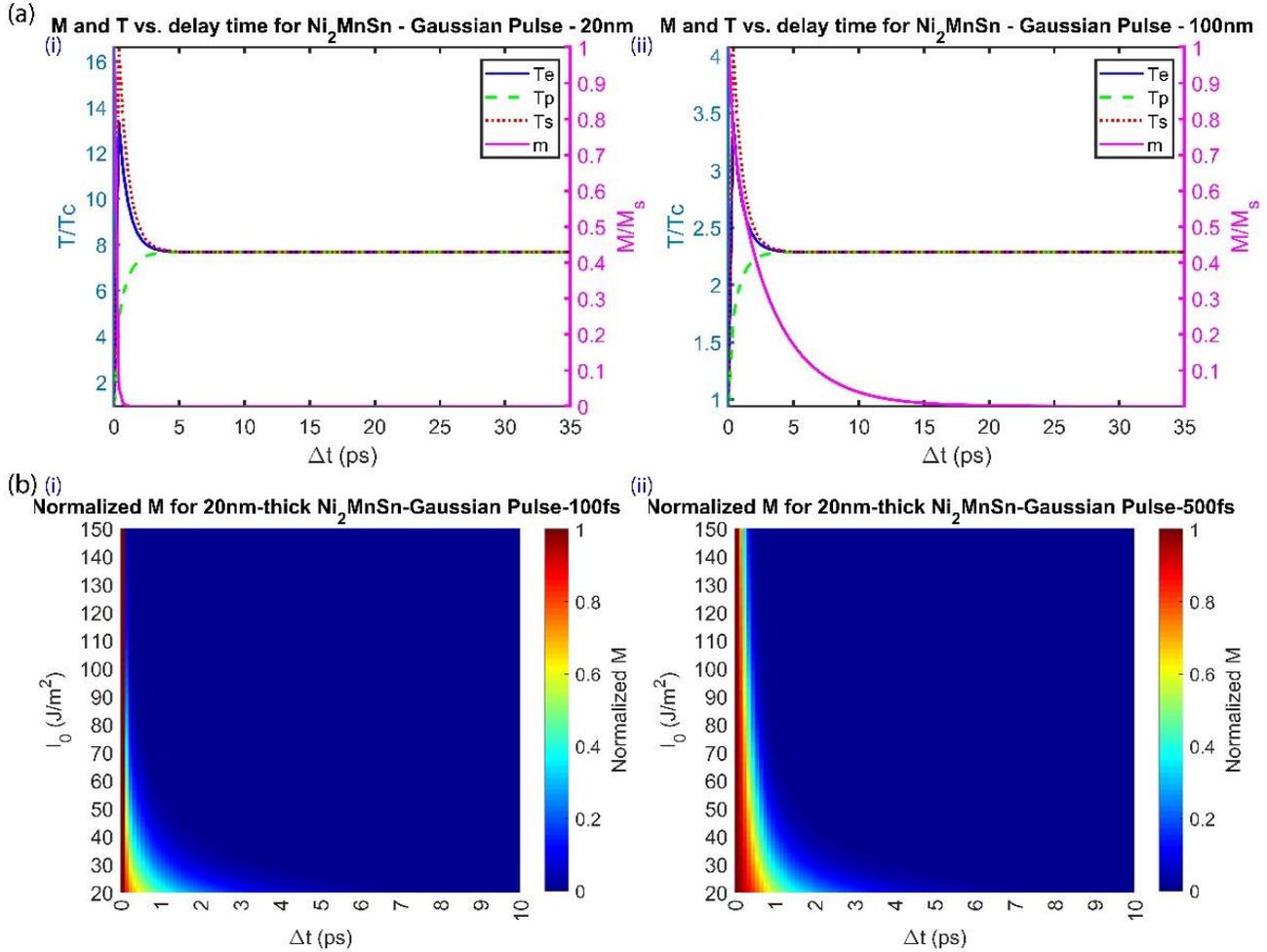

**FIG. 7. Effect of film thickness, fs laser fluence and pulse width on magnetization dynamics and lattice temperature of Ni$_2$MnSn magnetic thin film. a**(i) Transient electron ($T_e$), phonon ($T_p$), spin ($T_s$) temperatures, and normalized magnetization (m=|M$_z$|/M$_s$) of 20 nm and **a**(ii) 100 nm thick Ni$_2$MnSn illuminated with 100 fs laser pulse with 70 J·m$^{-2}$ fluence. **b**(i) Fluence dependence of magnetization of 20 nm thick Ni$_2$MnSn film illuminated with 100 fs and **b**(ii) 500 fs laser pulse.

### 4. THz spin wave spectra generated in type II films

Comparison of the spin wave spectra in 20 nm and 100 nm Gd and Ni$_2$MnSn thin films show that THz spin waves up to 8 THz could be generated in such thin films by manipulation of magnetization with 100 fs Gaussian laser pulse. One can tune the intensity and the bandwidth of the generated THz signal by changing the film thickness and laser pulse fluence. Laser pulses with sub-picosecond widths have minimal effect on the bandwidth and the intensity of the THz spin wave signal (see the spectra corresponding to 500 fs pulse widths in Supplemental Material). According to Fig. 8, the bandwidth and the intensity of the THz signal is higher in both 100 nm thick Gd and Ni$_2$MnSn thin films compared to 20 nm films since thicker films stay magnetized for longer. Hence, decreasing the laser fluence helps generate broadband THz spin waves with higher intensity.

Comparison of Fig. 5 and Fig. 8 shows that using magnetic metallic thin films which follow the type I dynamics results in generation of higher intensity and broader band THz spin waves, since type I metals preserve their magnetization for longer after interacting with the pulse.

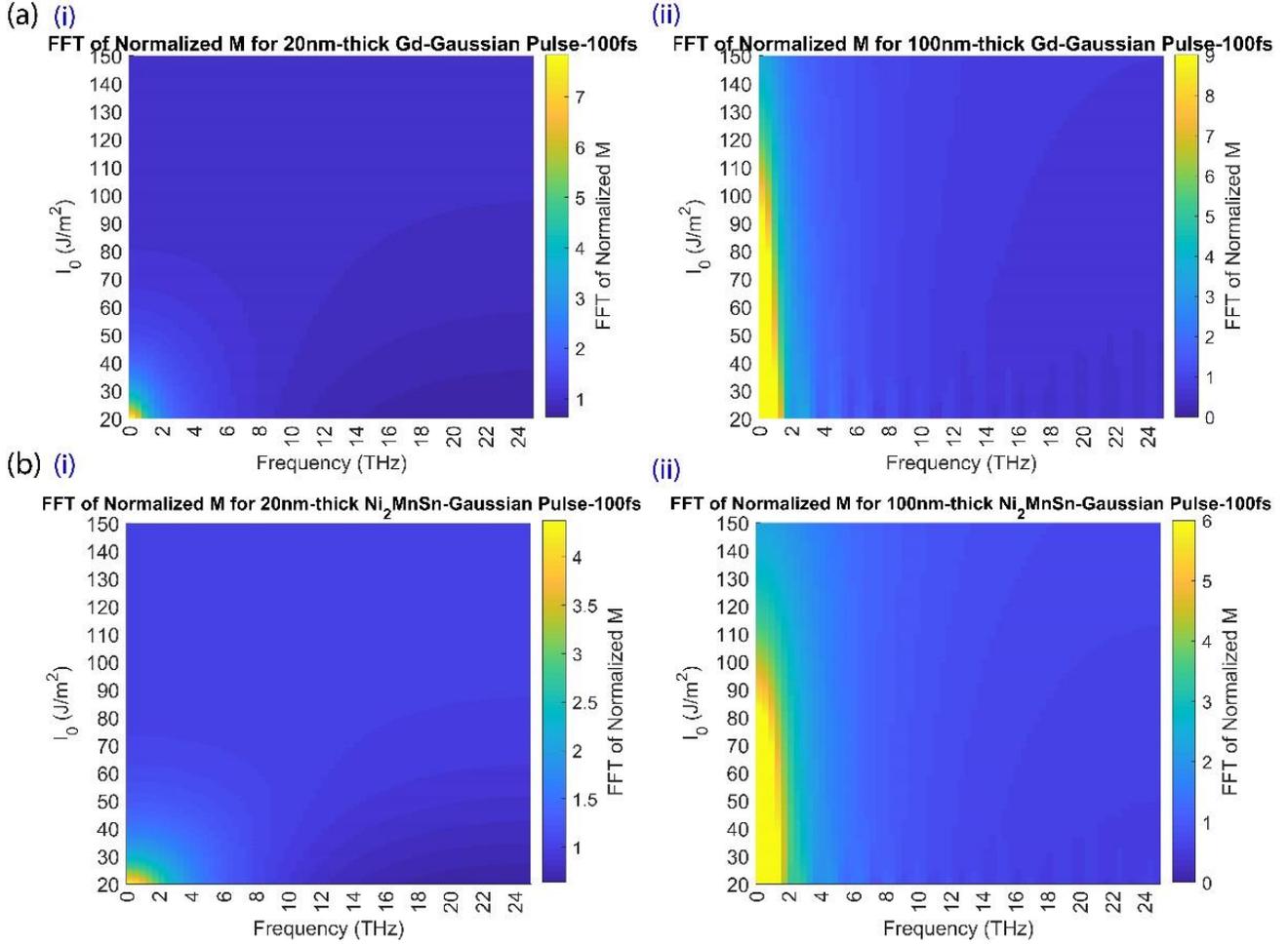

**FIG. 8. THz spin wave generation in type II metallic magnetic thin films.** Effect of laser fluence on the bandwidth and intensity of THz spin wave generated in **(a)** (i) 20nm and (ii) 100 nm thick Gd and **(b)** (i) 20nm and (ii) 100 nm thick $Ni_2MnSn$ films under illumination of 100 fs laser pulse.

### IV. CONCLUSION

We developed a model which explains the energy transfer from fs laser pulse to coupled electron, phonon, spin, and normalized magnetization in Fe, Co, Ni (type I films), Gd and $Ni_2MnSn$ (type II films). Our model and numerical solutions show that type I metallic thin films with Curie temperatures much higher than room temperature recover their magnetization. The films were first illuminated with Gaussian single laser pulse. Then the magnetization vector decreases briefly and reverts to its initial orientation within the next 130-200 fs. In the following 1-2 ps, the magnetization settles to its steady-state orientation. Increasing the laser pulse width and fluence increases the film response time and decreases its recovered fraction of magnetization. Type II films undergo thermal demagnetization gradually in about 35-50 ps after interaction with Gaussian laser pulse. Although higher laser fluences and pulse widths result in higher response times in these films, similar to type I films, increasing the thickness does not shorten the thermal demagnetization time. In thicker type I films (i.e. 100 nm), unlike type II films, the response time of magnetization is shorter. FFT of normalized magnetization dynamics show the broadband THz exchange-mediated spin wave generation up to 24 THz depending on laser pulse width, fluence and film thickness.

Our study suggests the need for optimizing the experimental parameters, such as laser pulse width and energy, film type and thickness for more energy efficient and faster response. Broadband THz spin waves generated with ultrafast all optical manipulation of magnetization could be utilized in many applications such as communications, imaging, security and sensing using THz spintronic time-domain spectroscopy (THz-sTDS).

## ACKNOWLEDGMENTS

M.C.O. acknowledges BAGEP 2017 Award, TÜBA-GEBİP Award by Turkish Academy of Sciences and TUBITAK Grant No. 117F416.

# Supplementary Information for "Ultrafast All Optical Magnetization Control for Broadband Terahertz Spin Wave Generation"

Saeedeh Mokarian Zanjani[1], Mehmet Cengiz Onbaşlı[1,2,*]

[1]Graduate School of Materials Science and Engineering, Koç University, Sarıyer, 34450 Istanbul, Turkey.
[2] Department of Electrical and Electronics Engineering, Koç University, Sarıyer, 34450 Istanbul, Turkey.
* Corresponding Author: monbasli@ku.edu.tr

**SUPPLEMENTAL MATERIAL**

### SUPPLEMENTAL NOTE 1

In the main manuscript, we solved four coupled differential equations, which describe the energy transfer between electron, phonon (lattice), spin baths and the normalized magnetization, after interaction with femtosecond (fs) Gaussian single laser pulse. Coefficients and material parameters used in solving the equations (1)-(4) in the main manuscript are shown in the Supplementary Table I. We have shown the magnetization dynamics and transient change of three temperature baths ($T_e$: electron, $T_p$: phonon/lattice, $T_s$: spin) for five different metallic thin films: Iron (Fe), Cobalt (Co), Nickel (Ni), Gadolinium (Gd), and Ni$_2$MnSn Heusler alloy. In addition, fast Fourier transform (FFT) results of the transient magnetization are plotted as the frequency-domain change of magnetization to illustrate the broadband THz spin wave generation. We applied our model to different thicknesses of the films illuminated with different laser pulse widths.

SUPP. TABLE I. Material parameters used in the extended M3TM model. $C_p$, $C_s$ are in J·m$^{-3}$·K$^{-1}$; $G_{ep}$, $G_{es}$, $G_{ps}$ are in W·m$^{-3}$·K$^{-1}$, $T_C$ is the Curie temperature (Kelvin), R: spin-flip ratio (s$^{-1}$), γ in J·m$^{-3}$·K$^{-2}$ [1-4].

| Film | $C_p \times 10^6$ | $C_s \times 10^6$ | $G_{ep} \times 10^{18}$ | $G_{es} \times 10^{18}$ | $G_{ps} \times 10^{18}$ | $T_C$ | $R \times 10^{12}$ | $\gamma = C_p/5T_C$ |
|---|---|---|---|---|---|---|---|---|
| Ni | 2.33 | 0.2 | 4.05 | 0.6 | 0.03 | 627 | 17.2 | 743.22 |
| Co | 2.07 | 0.2 | 4.05 | 0.6 | 0.03 | 1388 | 25.3 | 298.28 |
| Fe | 3.46 | 0.17 | 0.7 | 0.06 | 0.03 | 1043 | 1.86 | 663.47 |
| Gd | 1.78 | 0.2 | 0.21 | 0.6 | 0.03 | 297 | 0.092 | 1200 |
| Ni$_2$MnSn | 0.615 | 0.2 | 0.24 | 0.6 | 0.03 | 319 | 0.1 | 385.58 |

### SUPPLEMENTAL NOTE 2

This supplemental part is organized to present the calculated plots of sensitivity of the transient magnetization to the incoming laser pulse fluence and film thickness after illumination with 500 fs pulse width. Supp. Fig S1 and S2 show the sensitivity of the magnetization dynamics to the laser fluence and pulse width for 100 nm thick type I and type II thin films, respectively. Supp. Fig. S3 and S4 show the sensitivity of the spectral change of magnetization to the laser fluence and film thickness for 500 fs laser pulse width for type I and type II thin films, respectively.

### SUPPLEMENTAL NOTE 3

The fluence dependence of magnetization dynamics of 20 nm films are presented in the main manuscript. The sensitivity of magnetization to the fluence and pulse width for 100 nm thick Fe, Co, and Ni films are shown in Supp. Fig. S1(a, b, c), respectively. The magnetization dynamics have been calculated for fluences from 20 to 150 J·m$^{-2}$. Illuminated with both 100 fs (Supp. Fig. S1. a(i), b(i), c(i)) and 500 fs (Supp. Fig. S1. a(ii), b(ii), c(ii)), both magnetization switching and recovery times of 100 nm-thick films increase with increasing laser fluence. The recovered magnetization fraction, however, decreases with increasing laser fluence. There is a threshold fluence above which the films are thermally demagnetized. This threshold fluence depends on the film type, thickness and pulse width. Increasing the pulse width increases the response time of the films.

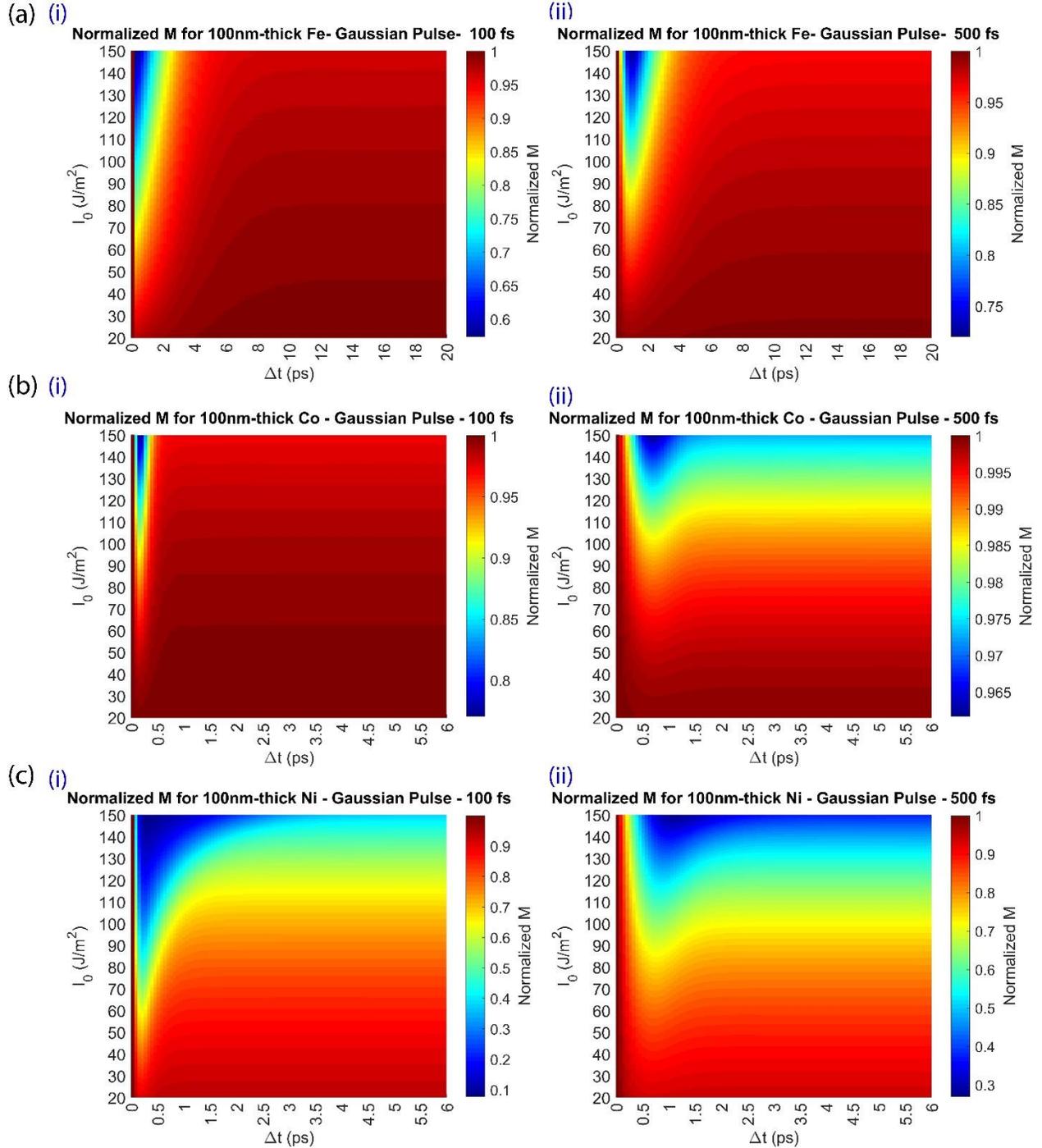

**SUPP. FIG. S1. Effect of fs laser fluence and pulse width on magnetization dynamics and lattice temperature of type I magnetic thin films.** Sensitivity of transient magnetization ($m=|M_z|/M_s$) of 100 nm thick (**a**) Fe, (**b**) Co, and (**c**) Ni under illumination of (i) 100 fs and (ii) 500 fs laser pulse widths.

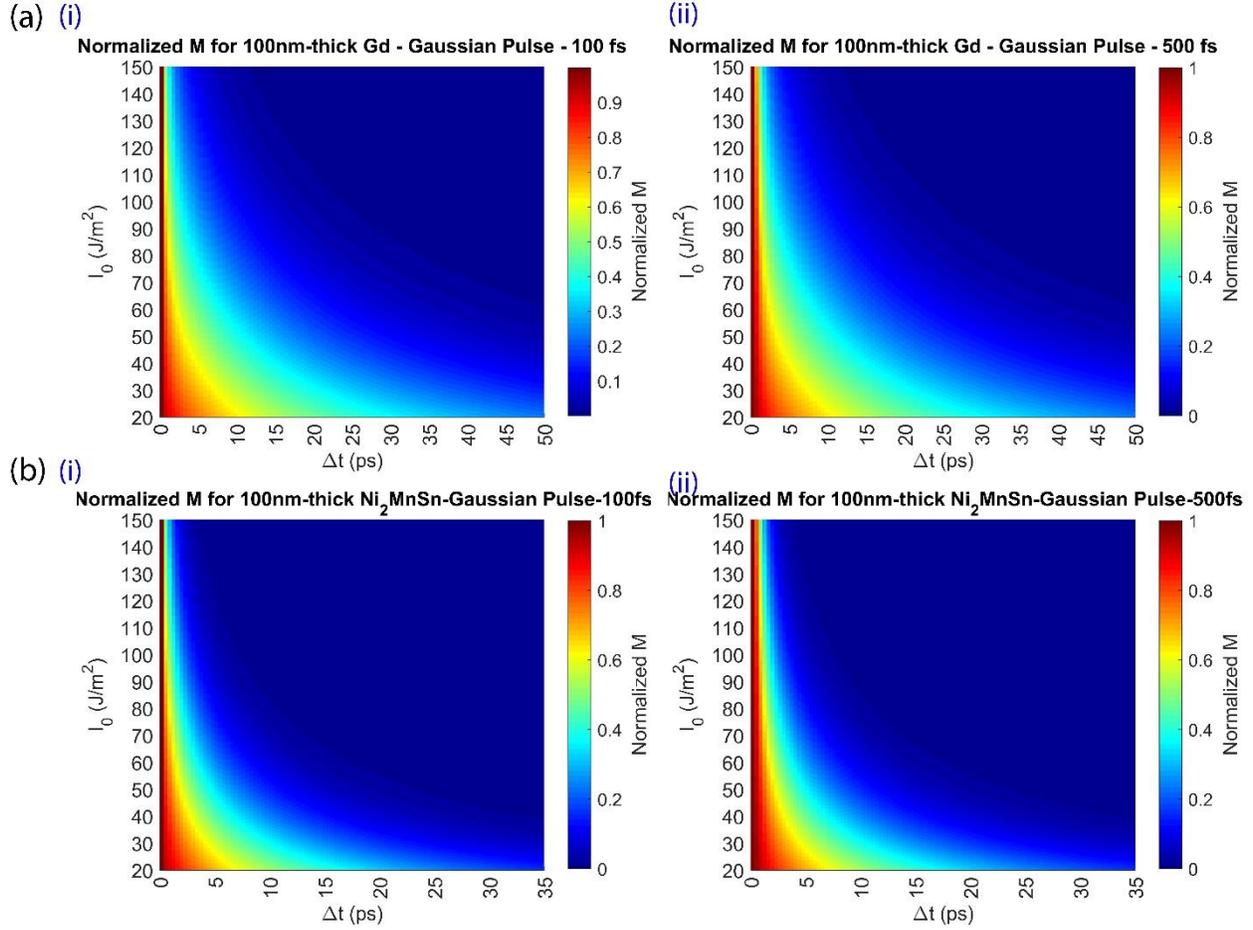

**SUPP. FIG. S2. Effect of fs laser fluence and pulse width on magnetization dynamics and lattice temperature of magnetic type II thin films.** Sensitivity of transient magnetization (m=|$M_z$|/$M_s$) of 100 nm thick (**a**) Gd, (**b**) $Ni_2MnSn$ under illumination of (i) 100 fs and (ii) 500 fs laser pulse widths.

Supp. Fig. S2(a) and S2(b) show the dependence of magnetization dynamics to the pulse fluence and pulse width for 100 nm thick Gd and $Ni_2MnSn$ films, respectively. Type II films, unlike type I, undergo total thermal demagnetization and their moments cannot be recovered. Similar to type I dynamics, increasing the laser fluence and pulse duration increases the response time in both type II films. On the other hand, excitation of the films with smaller laser fluences keeps the films magnetized for longer times. Unlike type I films, the response time increases in type II films by increasing the film thickness. Besides, increasing the incoming laser pulse width increases the response time in both films, as shown in Supp. Fig. S2(a(ii)) and (b(ii)).

### SUPPLEMENTAL NOTE 4

Supp. Fig. S3 shows the FFT of the fluence dependence of temporal magnetization dynamics for Fe, Co and Ni. The spectra for films under 100 fs laser pulse illumination are presented in the main manuscript. Supp. Fig. S3 shows the spectra for Fe, Co, and Ni films each illuminated with single 500 fs Gaussian laser pulses. Spin waves up to 10 THz could be generated. The spin wave intensity is higher when fs laser fluence is lower. Increasing film thickness to 100 nm increases the THz spin wave intensity slightly, and the allowed threshold fluence for THz spin wave generation. From the spectra of Fe and Co, one can infer that the THz spin wave intensity is slightly higher compared to Ni, since these two metals have higher $T_C$ and recovered magnetization fractions (> 95%).

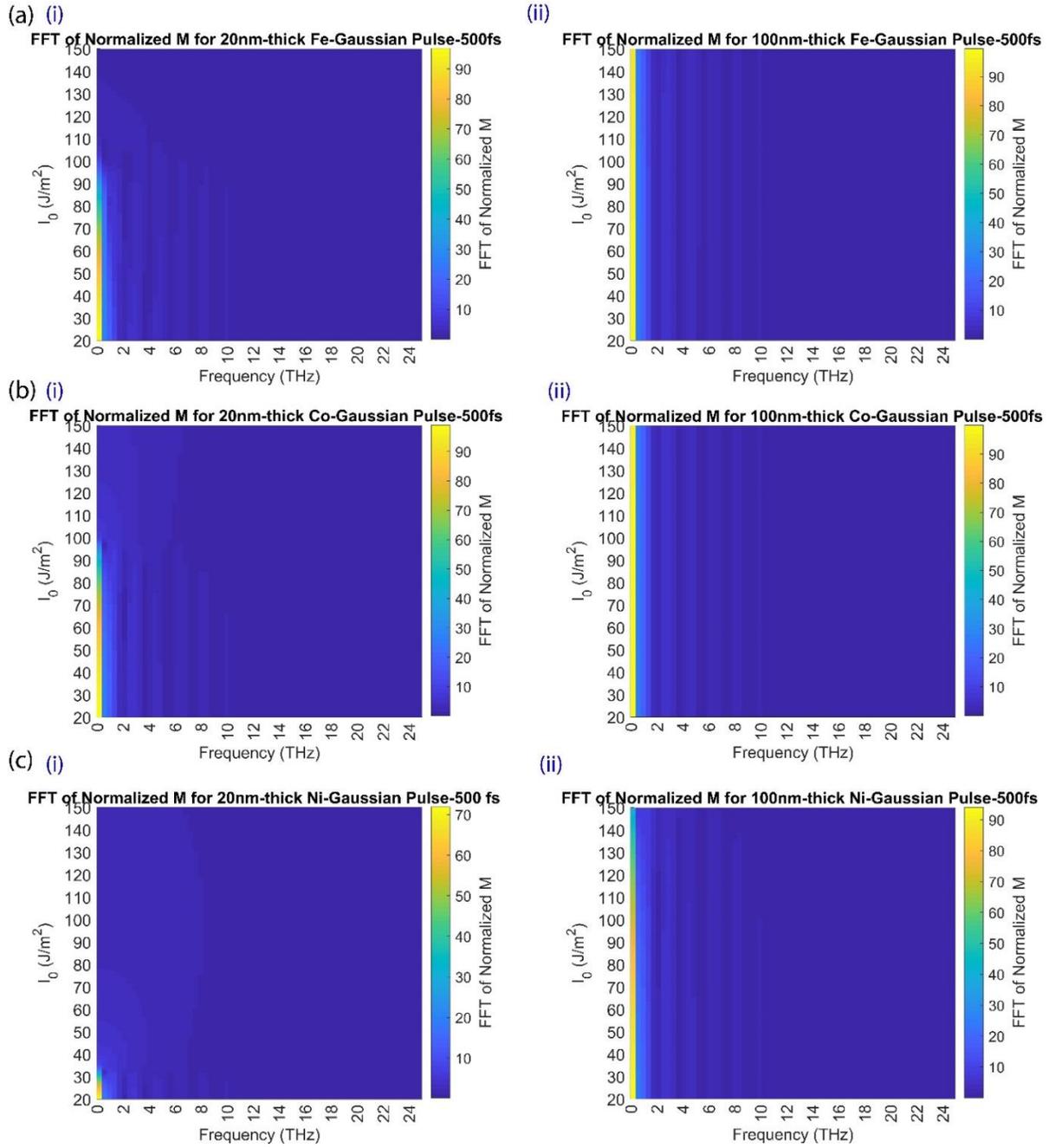

**SUPP. FIG. S3. THz spin wave generation in type I metallic magnetic thin films.** Effect of laser fluence on the bandwidth and intensity of THz spin waves generated in **a**(i) 20nm and **a**(ii) 100 nm thick Fe, **b**(i) 20nm and **a**(ii) 100 nm thick Co, and **c**(i) 20nm and **c**(ii) 100 nm thick Ni films each illuminated under 500 fs single laser pulse.

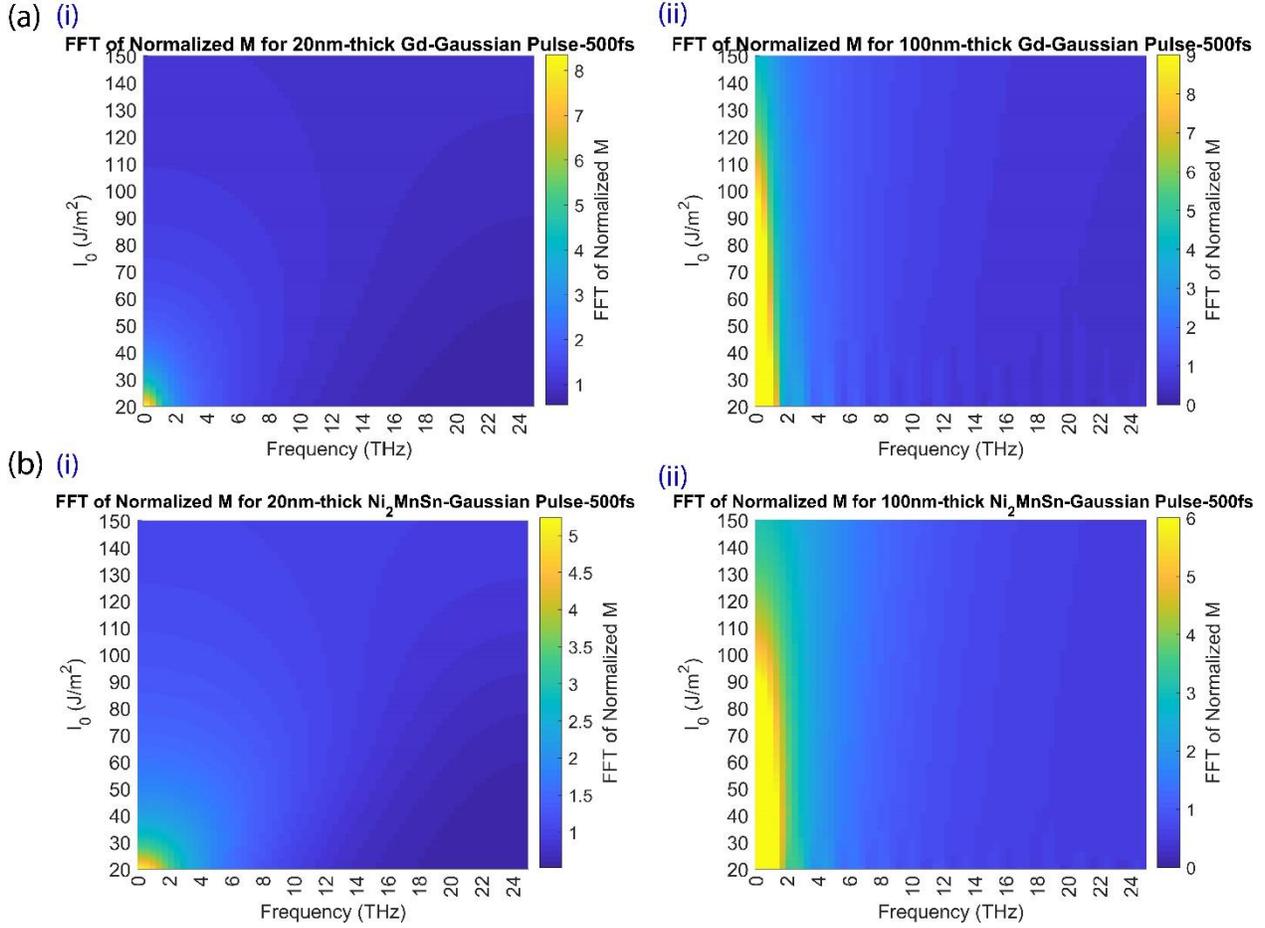

**SUPP. FIG. S4. THz spin wave generation in type II metallic magnetic thin films.** Effect of laser fluence on the bandwidth and intensity of THz spin wave generated in **a**(i) 20nm and **a**(ii) 100 nm thick Gd, **b**(i) 20nm and **b**(ii) 100 nm thick $Ni_2MnSn$ films under illumination of 500 fs single laser pulse.

Comparison of the spin wave spectra in 20 nm and 100 nm Gd and $Ni_2MnSn$ thin films show that THz spin waves up to 8 THz could be generated in such thin films by manipulation of magnetization with 500 fs Gaussian laser pulse. One can tune the intensity and the bandwidth of the generated THz spin waves by changing the film thickness and laser pulse fluence. Laser pulses with sub-picosecond widths have minimal effect on the bandwidth and the intensity of the THz spin wave signal. According to Supp. Fig. S4, the bandwidth and the intensity of the THz spin waves is higher in both 100 nm Gd and $Ni_2MnSn$ thin films compared to 20 nm films since thicker films stay magnetized longer. Hence, decreasing the laser fluence helps generate broadband THz spin waves with higher intensity. Comparison of Supp. Fig. S4 and S3 shows that using type I films results in generation of higher intensity and broader band THz spin waves, since type I metals preserve their magnetization for longer after interacting with the pulse.